\documentclass[pss]{wiley2sp} % provides pss two-column style
\usepackage{caption}
\usepackage{caption3}
\begin{document}

% Title of the article
\title{Cleaning graphene: comparing heat treatments in air and in vacuum}

% Abbreviated title for the page headers
\titlerunning{Short title }

% Authors
\author{%
  Mukesh Tripathi,
  Andreas Mittelberger,
  Kimmo Mustonen,
  Clemens Mangler,
  Jani Kotakoski,
  Jannik C. Meyer and
  Toma Susi\textsuperscript{\Ast}}

% Abbreviated list of authors for the page headers
\authorrunning{Tripathi et al.}

%E-mail-address of corresponding author
\mail{e-mail
  \textsf{toma.susi@univie.ac.at}, Phone:
  +43-01-4277-72855}

% author's affiliations/addresses
\institute{%
  Faculty of Physics, University of Vienna, Boltzmanngasse 5, Vienna, 1090, Austria\\
 }

\received{XXXX, revised XXXX, accepted XXXX} % do not change, will be filled in by the publisher
\published{XXXX} % do not change, will be filled in by the publisher

% Please select about four verbal keywords for your manuscript.

\abstract{
\abstcol{
  Surface impurities and contamination often seriously degrade the properties of two-dimensional materials such as graphene. To remove contamination, thermal annealing is commonly used. We present a comparative analysis of annealing treatments in air and in vacuum, both \textit{ex situ} and \textit{"pre-situ"}, where an ultra-high vacuum treatment chamber is directly connected to an aberration-corrected scanning transmission electron microscope. While \textit{ex situ} treatments do remove contamination, it is challenging to obtain atomically clean surfaces after ambient transfer. However, \textit{pre-situ} cleaning with radiative or laser heating appears reliable and well suited to clean graphene without undue damage to its structure.}}
\titlefigure[]{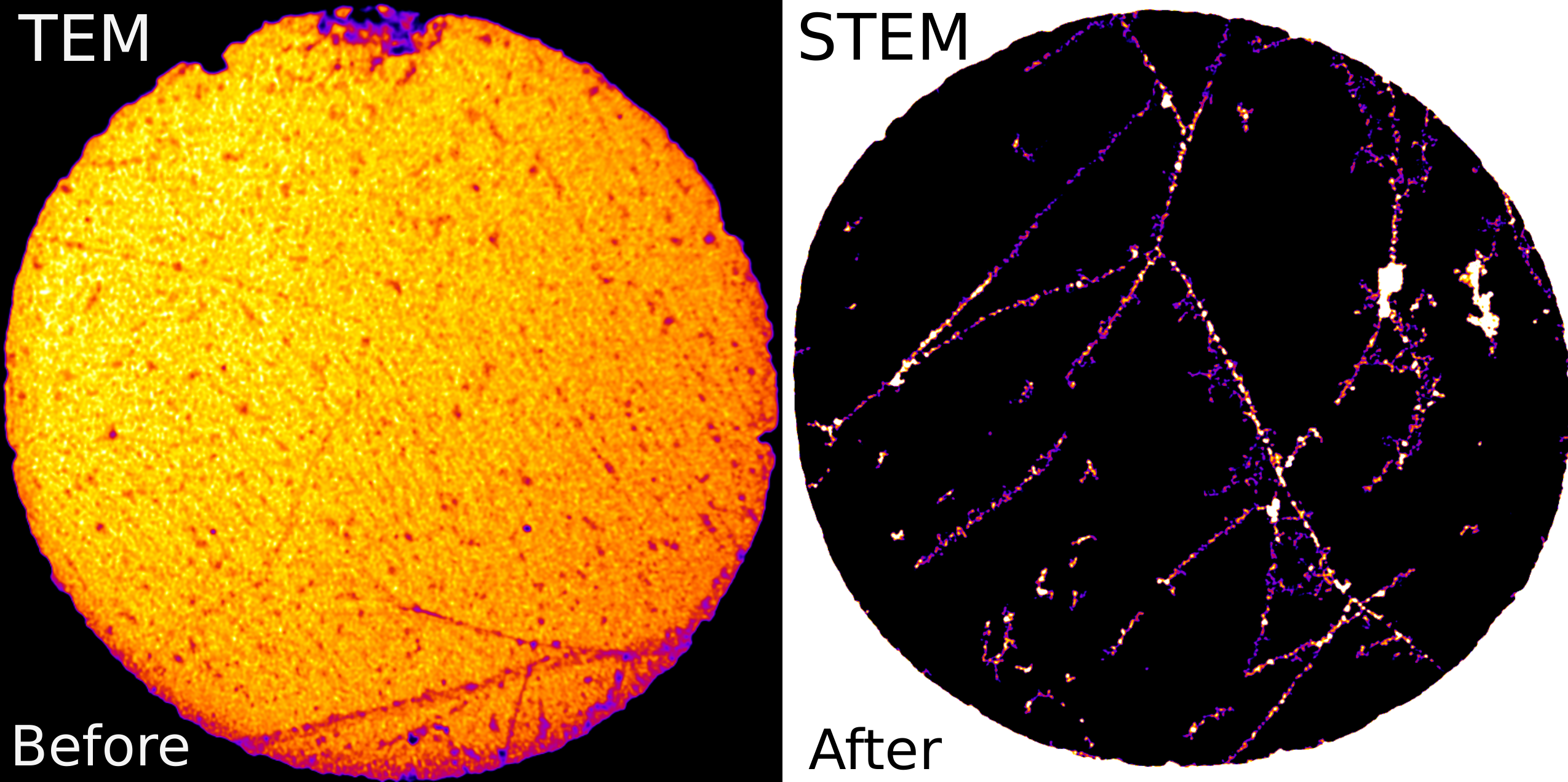}
\titlefigurecaption{\textit{Pre-situ} annealing of typical dirty graphene samples yields atomically clean areas several hundred nm$^2$ in size.}

\maketitle

\section{Introduction}
Graphene~\cite{Novoselov20004Science} has attracted considerable attention due to its excellent intrinsic properties, leading to many potential applications including DNA translocation~\cite{Merchant2010nanolett.}, nanoelectronic devices~\cite{Becerril2008ACSnano}, and sensors~\cite{Schedin2007Natmater}. Chemical vapor deposition (CVD) allows large area graphene to be synthesized scalably and in high-yield on transition metal surfaces, from which polymers such as poly methyl methacrylate (PMMA) are used to transfer it onto target substrates~\cite{Li2009Science}. To dissolve PMMA after transfer, organic solvents like acetone, chloroform and acetic acid are commonly used ~\cite{Cheng2011Nanoletters,Her2013PhylettA}. However, none of these solvents are able to completely dissolve PMMA, and a thin layer of polymeric residues are left absorbed on the surfaces~\cite{Lin2011ACSNANO}. This is a major drawback of polymer-assisted transfer and can degrade the electronic properties of graphene by introducing unintentional doping and charge impurity scattering~\cite{Pirkle2011APL}. In addition, hydrocarbon impurities are directly absorbed from the atmosphere onto the surface, and mobile contamination may be pinned into place by the electron beam~\cite{Meyer08APL}. This makes atomic level characterization by electron microscopy and electron energy loss spectroscopy~\cite{Susi172DM} difficult, not to mention more ambitious goals such as single-atom manipulation~\cite{Susi15FWF,Susi17UM}.

To clean graphene, several methods have been reported. Conventional thermal annealing is optimized by varying the treatment temperature in air~\cite{Xie2015Carbon,Wang2017Chem.ofMater.}, in vacuum~\cite{Cheng2011Nanoletters,Pirkle2011APL,Lin2012Nanolett.,Ni2010Jour.oframanspectroscopy} and in gas environments such as Ar/H$_2$~\cite{W.choi2015IEEE,Ahn16Mater.Express}, CO$_2$~\cite{Gong2013Journofphychem} or N$_2$~\cite{Jang2013Nanotech}. Moreover, vacuum annealing at higher temperature for shorter times, i.e. rapid-thermal annealing ~\cite{Jang2013Nanotech}, has been successfully used to remove surface contamination. Several other approaches such as dry-cleaning with activated carbon~\cite{Algara-Siller14APL}, wet chemical treatment using chloroform~\cite{Cheng2011Nanoletters}, and deposition of metal catalyst and subsequent annealing~\cite{Longchamp2013Jourofvacscience} have been studied. However, adsorbents or chemicals also leave residues, and depositing metal will affect transport and other properties. Non-chemical routes such as mechanical cleaning using contact mode atomic force microscope~\cite{Goossens2012APL} or plasma treatment~\cite{Ferrah2016surfandinterfaceanaly.} have been employed, but have limited ability to remove contamination over large areas.

In this work, we analyze and compare the effectiveness of heat treatments in air and in vacuum to clean graphene. We investigate its relative cleanliness after \textit{ex situ} annealing in air on a hot plate or in a vacuum chamber. We further demonstrate a new, effective and reliable cleaning approach using black body radiative or laser-induced heating in vacuum. In this "\textit{pre-situ} cleaning", the sample is annealed in the same vacuum system as the characterization equipment, to which it is transferred without exposure to the ambient. While this is a standard technique for surface science, it has until now not been possible to combine it with electron microscopy. To study the effectiveness of methods used, our samples were characterized using low acceleration voltage transmission electron microscopy (TEM) and atomic resolution aberration-corrected scanning transmission electron microscopy (STEM). 

We find that while \textit{ex situ} treatments do remove contamination, when effective they also cause significant damage. Only with the \textit{pre-situ} method was it possible achieve large areas of atomically clean graphene.

\section{Experimental}
Commercially available CVD-grown monolayer graphene suspended on Quantifoil TEM grids from Graphenea Inc. was used for the experiments. All \textit{ex situ} samples were characterized using a bench-top low acceleration voltage transmission electron microscope (LVEM5, 5 kV). Selected \textit{ex situ} and all \textit{pre-situ} samples were characterized at high resolution using the aberration-corrected scanning transmission electron microscope Nion UltraSTEM100 operated at 60~kV (with a standard 12~h 130~$^{\circ}$C vacuum bake before insertion into the microscope, apart from the radiatively heated samples inserted via a separate airlock). All presented images have been treated with a Gaussian filter and colored with the ImageJ lookup table "fire" to highlight the relevant details.

We used two \textit{ex situ} cleaning techniques: air and vacuum annealing. In air, samples were heated on a hot plate between 300--500~$^{\circ}$C for times ranging from 15~min to 1~h. Vacuum annealing was carried out in a vacuum evaporator (Korvus Technology) at a pressure of 10$^{-6}$~Torr. TEM grids were inserted into the vacuum chamber in a ceramic bucket wrapped with a resistive coil and a thermocouple placed inside to measure the temperature.

For \textit{pre-situ} annealing, we likewise used two techniques: laser annealing and radiative heating. For laser annealing, a high power diode laser (tunable up to 6~W) was aimed through a viewport at the sample held in the parked pneumatic transfer arm. The samples were iteratively treated with increasing laser power until cleaning was observed, leading to good results with 600~mW (10~\% duty cycle) for 2 min. The laser spot was $\sim$1~mm$^2$ in size and the distance between the laser source and sample was $\sim$40~cm. The power must be carefully controlled since at higher power the laser will destroy the sample. 

Radiative heating was effected by a tungsten (W) wire that can be resistively heated to very high temperatures, mounted in a vacuum chamber attached to the microscope. Distance between the wire and sample was $\sim$2--3~mm and the treatment time was 15 min. Again, the wire power was iteratively increased until cleaning was observed, yielding good results for a current of 7~A, corresponding to a thermal power of 64~W and a wire temperature of $\sim$1750~K. The vacuum level for both \textit{pre-situ} methods was $\sim$10$^{-8}$~Torr. For imaging, the samples were transferred into the Nion UltraSTEM without exposure to air.

\section{Results and discussion}
\begin{figure}[!t]%
\includegraphics*[width=.48\textwidth]{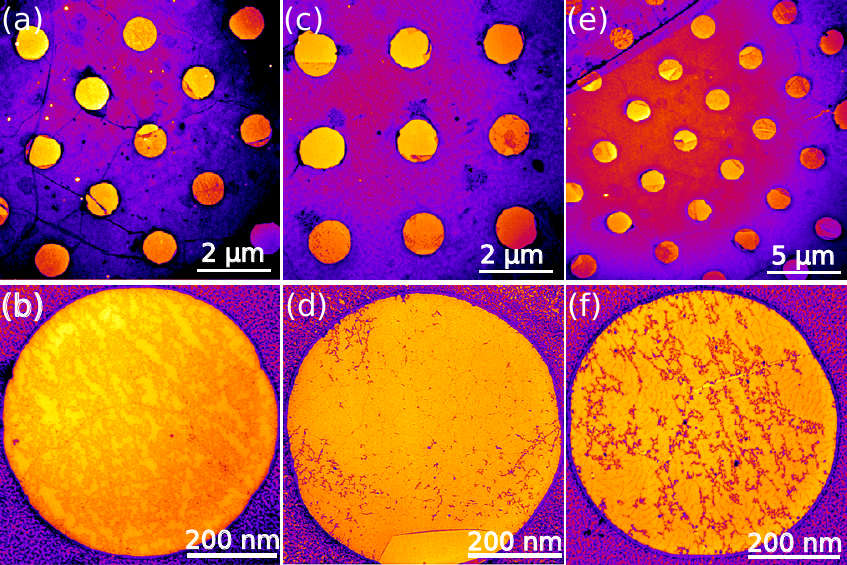}
\caption{TEM images of graphene after heat treatment in air. (a) Overview and (b) magnified view after annealing at 400~$^{\circ}$C for 1 h. (c) Overview and (d) magnified view after 450~$^{\circ}$C for 30 min. (e) Overview and (f) magnified view after 500~$^{\circ}$C for 15 min.}
\label{airannealing}
\end{figure}

Figure 1 shows low voltage TEM images of suspended monolayer graphene after annealing in air at temperatures between 400--500~$^{\circ}$C (treatment at lower temperatures does not yield larger clean areas, even if contamination layers are thinner). After air treatment at 400~$^{\circ}$C for 1 h, structural damage starts to emerge, but residues have not been much affected as shown in Fig.~\ref{airannealing}a and b. By increasing the temperature to 450~$^{\circ}$C for 30 min, tearing of graphene sheets becomes more frequent and the concentration of impurities is reduced as illustrated in Fig.~\ref{airannealing}c and d. However, significant contamination still remains. At 500~$^{\circ}$C for 15 min, crack formation is evident almost everywhere on the sample, while the density of residues decreases further as shown in Fig.~\ref{airannealing}e and f. At the same time, some contamination regions appear to be thicker after the treatment. A two-step treatment of washing the sample with aqueous acetonitrile and baking in air did not show additional effect. Thus, air annealing at high temperatures does help in removing residues, but severe damage occurs in the suspended graphene regions, presumably assisted by the etching of grain boundaries.

\begin{figure}[t]%
\includegraphics*[width=0.48\textwidth]{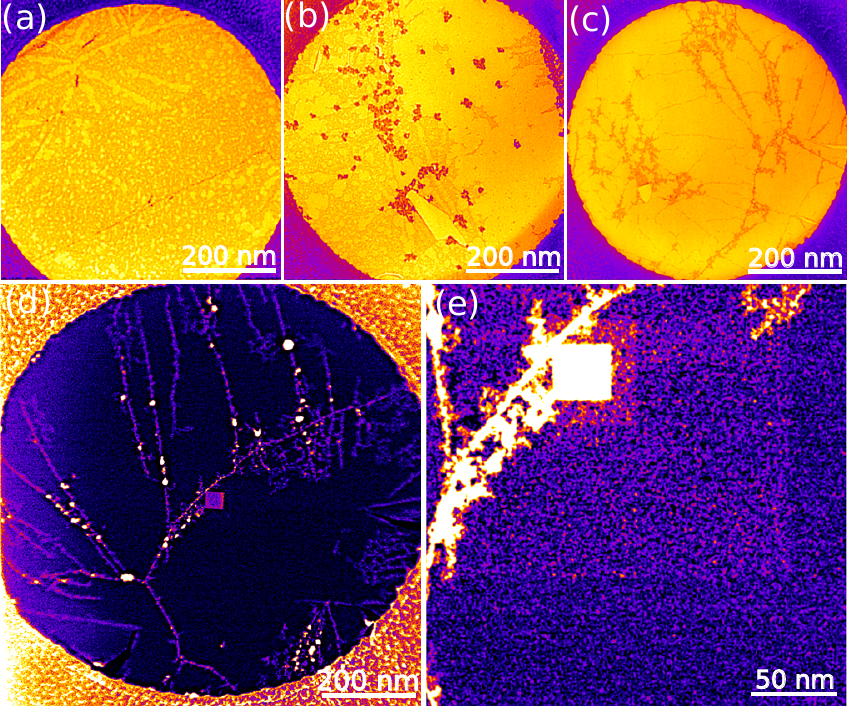}%
  \caption[]{%
TEM and STEM images of graphene after vacuum annealing. (a) 600 $^{\circ}$C for 30 min, (b) 650 $^{\circ}$C for 15 min, (c) 750 $^{\circ}$C for 3 min. After annealing at 750 $^{\circ}$C, this sample was transferred in ambient to the Nion UltraSTEM. (d) Low magnification STEM image and (e) magnified view of a clean-looking area, revealing a layer of contamination still covers the surface, and more rapidly accumulates under the beam (bright squares).}\label{vacuumannealing} 
\end{figure}

In vacuum, graphene can withstand significantly higher temperatures. Fig.~\ref{vacuumannealing} shows TEM and STEM images of graphene annealed between 600--750~$^{\circ}$C (heated at a rate of 10 $^{\circ}$C/min and cooled to room temperature in N$_2$) and subsequently characterized for cleanliness. The TEM images in Fig. 2a of a sample heated to 600~$^{\circ}$C show that contaminants are covering the surface, with small clean spots no larger than a few tens of nm$^2$. After thermal treatment at 650~$^{\circ}$C for 15~min, surface contamination was reduced (Fig.~\ref{vacuumannealing}b). However, long treatments at high temperature start to cause crack formation even in vacuum. We further increased the annealing temperature to 750 $^{\circ}$C but reduced the time to only 3 minutes, and observed that many contaminants had been removed (Fig.~\ref{vacuumannealing}c). We also found apparently almost fully clean areas, apart from some remaining chains of impurities as shown in Fig.~\ref{vacuumannealing}c. However, even this short treatment time resulted in severe tearing of the suspended graphene. While etching should be suppressed in vacuum, it may be that mismatch in the thermal expansion coefficients of graphene and the substrate causes severe mechanical stress that leads to the tearing.

To verify the cleaning, we imaged the 750~$^{\circ}$C sample at higher resolution in the STEM. The medium angular annular dark field (MAADF) image of Fig.~\ref{vacuumannealing}d shows the large clean-looking areas, and some chain-like impurity patterns. Since contrast in annular dark field (ADF) STEM is directly proportional to the atomic number and the number of atoms in the beam path~\cite{Krivanek2010nature}, the bright spots are possibly heavier elements such as gold particles from the gold support grid that have become mobile at high temperature. However, at higher magnifications, we observed that a thin layer of contamination is still covering the regions that appear clean at lower resolution. Furthermore, the square bright areas in Fig.~\ref{vacuumannealing}e were caused by mobile contamination pinned onto the surface by the electron beam. These findings may be explained by the highly lipophilic nature of graphene: a thin layer of contamination quickly adsorbs on the surface when graphene is exposed to the ambient ~\cite{Booth2008Nanoletters}. Alternatively, the contaminants may not be desorbed by the treatments, but merely swept aside into larger aggregates, only to diffuse back afterwards.

\begin{figure}[!hb]%
\includegraphics*[width=0.48\textwidth]{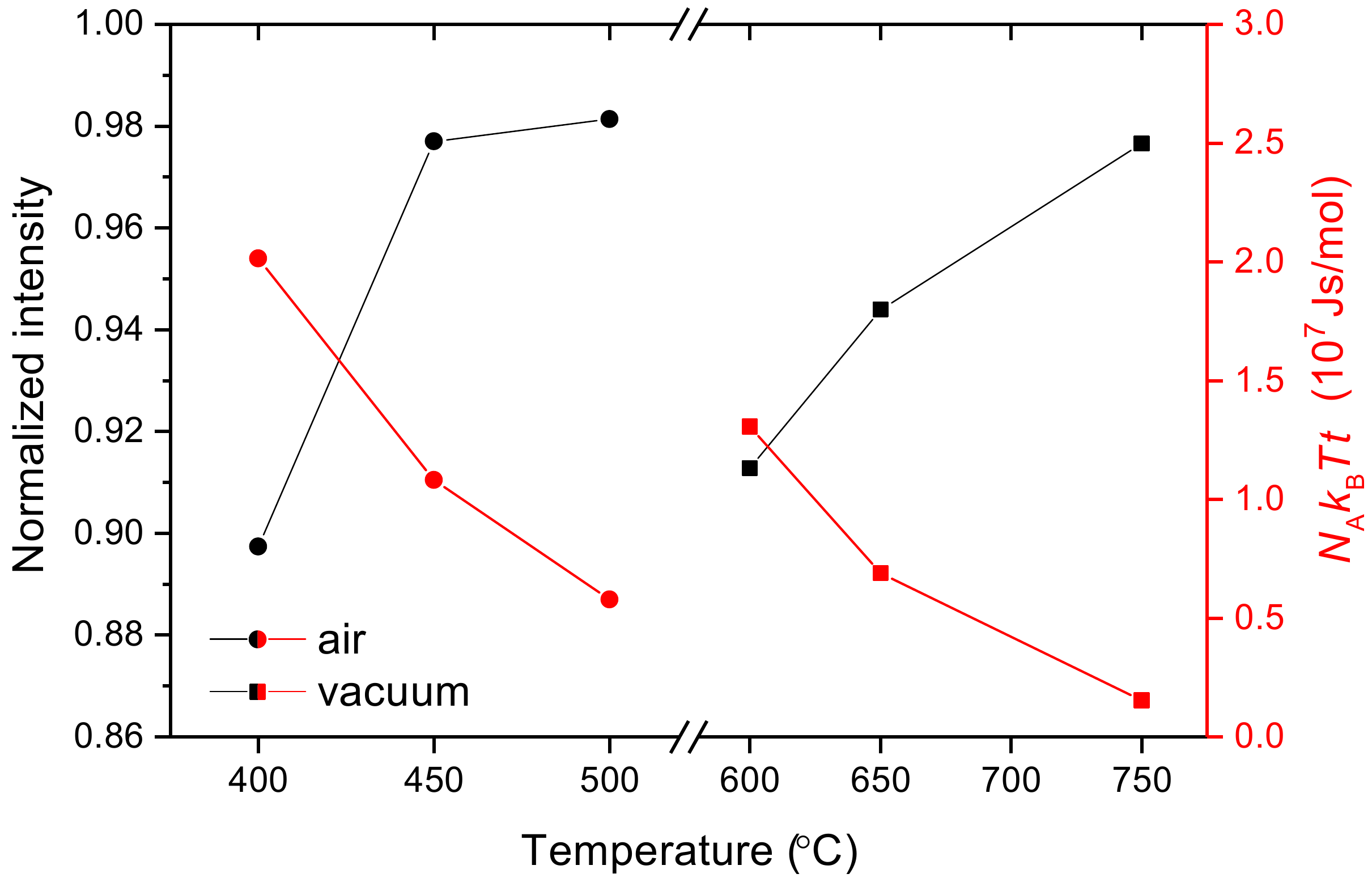}
\caption{%
Quantifying the effect of treatments in air and in vacuum. Left axis shows the normalized integrated intensity as a function of annealing temperature (for untreated samples, a typical value was $\sim$0.61), and right axis the thermal energy per mole multiplied by the treatment time (Eq.~\ref{eq1}). As the temperature increases, the integrated intensity approaches the vacuum level, corresponding to the reduction of impurities. Higher temperature treatments are more effective despite shorter treatment times.}
\label{cleaningplot}
\end{figure}

To quantify the effect of cleaning, in Fig.~\ref{cleaningplot} we plot the integrated intensity measured over several hundred nm$^2$ of graphene (normalized by the vacuum level to account for differences in beam focusing) for air and vacuum annealing at different temperatures. For both treatments, the integrated intensity approaches unity with increasing temperature, indicating a decrease of impurity concentration as contaminants on the surface diffuse away or are desorbed. Since we used different treatment times at different temperatures, we also calculated the time integral of the thermal energy per mole ("thermal action"), defined as

\begin{equation}
\label{eq1}
S_{th} = N_\mathrm{A} k_\mathrm{B} T t,
\end{equation}
where $N_\mathrm{A}$ is the Avogadro constant, $k_\mathrm{B}$ the Boltzmann constant, $T$ the temperature in Kelvin, and $t$ the treatment time. From the plot in Fig.~\ref{cleaningplot} we see that relatively shorter treatments are required at higher temperature for the same or even better cleaning effect. This corroborates the effectiveness of rapid-thermal annealing.

\begin{figure}[t]%
\includegraphics*[width=1\linewidth]{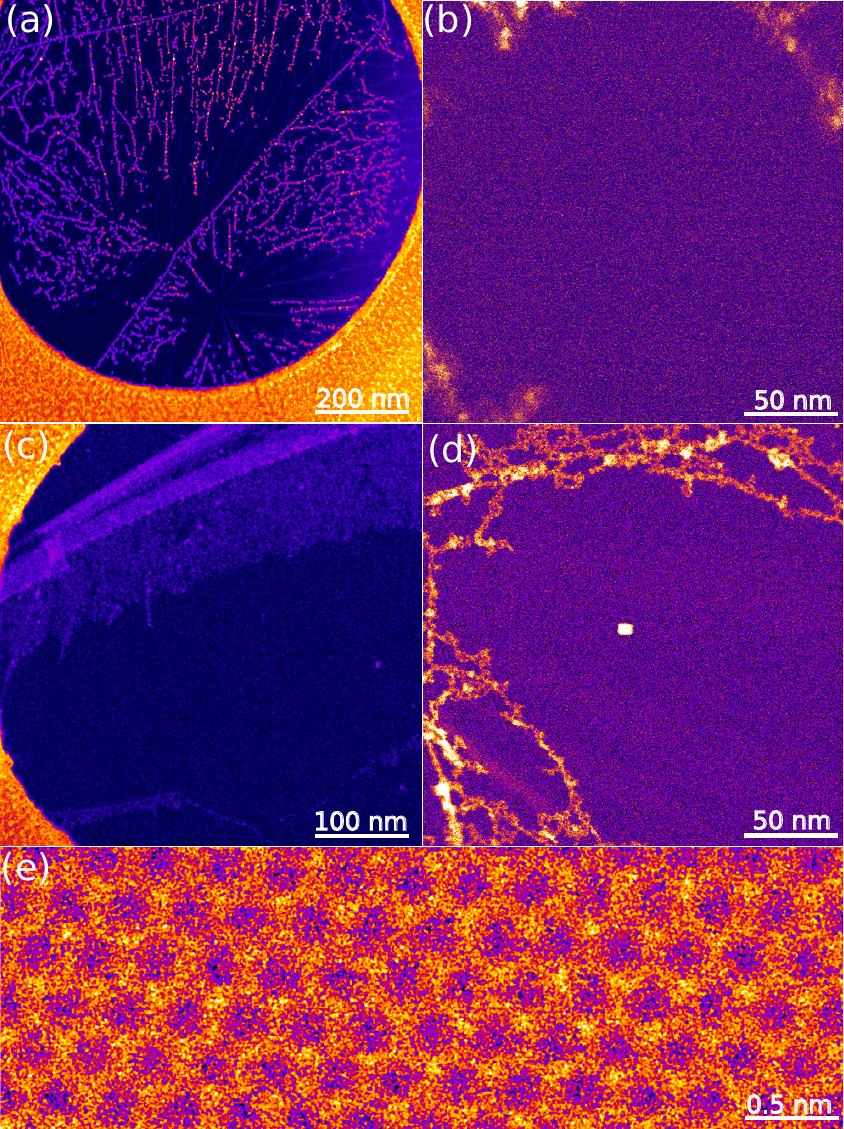}
\caption{%
STEM images showing cleaned graphene after \textit{pre-situ} annealing. (a) Low and (b) intermediate magnification images after radiative heating, and (c) low and (d) intermediate magnification images after laser-induced heating in vacuum. Panel (e) shows an example of the atomically resolved and clean graphene lattice (the non-ideal imaging conditions presumably resulted  from residual heat from the treatment).}
\label{presitu}
\end{figure}

To clean graphene using \textit{pre-situ} annealing in a custom-built vacuum chamber attached to column of the STEM, we made use of both radiative energy transfer from a resistively heated W wire and from a high power laser aimed at the sample. In the case of radiative heating, current and voltage were controlled using a lab power supply, and in both cases the sample was transferred for observation without breaking the vacuum. The MAADF images in Fig.~\ref{presitu}a and b show clean graphene after W wire heating. Surface contaminants is greatly reduced and large uniformly clean graphene regions are obtained as shown in Fig.~\ref{presitu}b. Results of the laser cleaning are shown in Fig.~\ref{presitu}c and d. Contaminants are mostly eliminated by the laser treatment, resulting in atomically clean areas of several hundred nm$^2$ (the MAADF image in Fig.~\ref{presitu}e shows an example of the atomically clean lattice). Interestingly, while we observed mobile contamination pinning under the beam, in most cases this occurred only when the field of view contained pre-existing contamination or other defects.

\section{Conclusions}
In conclusion, we have compared heat treatments to clean graphene in air and in vacuum. We clearly show that air annealing is not a good method: contamination remains on the surface, and severe damage occurs at higher temperatures where the treatment is more effective. Annealing at higher temperatures in vacuum is more effective in removing surface contaminants, but some seem to readsorb upon exposure to an air ambient. This issue can be overcome with \textit{pre-situ} annealing via radiative or laser-induced heating in the same vacuum system as the electron microscope. These methods appear to be reliable and controllable for cleaning graphene and potentially other 2D crystals. However, caution must be taken in selecting the treatment time and the laser or thermal power to avoid destroying the sample. With optimal parameters, large areas of atomically clean graphene can be easily obtained. 

\begin{acknowledgement}
M.T. and T.S. acknowledge funding by the Austrian Science Fund (FWF) via project P 28322-N36 and J.K. by the Wiener Wissenschafts\mbox{-,} Forschungs- und Technologiefonds (WWTF) via project MA14-009. A.M., K.M., C.M., and J.C.M. acknowledge funding from the European Research Council (ERC) Grant No. 336453-PICOMAT. K.M. acknowledges financial support from the Finnish Foundations’ Post Doc Pool.
\end{acknowledgement}

\providecommand{\WileyBibTextsc}{}
\let\textsc\WileyBibTextsc
\providecommand{\othercit}{}
\providecommand{\jr}[1]{#1}
\providecommand{\etal}{~et~al.}

	\end{document}